\pdfoutput=1

\documentclass[11pt]{article}

\usepackage[utf8]{inputenc}
\usepackage[T1]{fontenc}
\usepackage{lmodern}
\usepackage{microtype}
\usepackage[margin=1in]{geometry}

\usepackage{graphicx}
\usepackage{subcaption}
\usepackage{amsmath}
\usepackage{amsfonts}
\usepackage{amssymb}
\usepackage{booktabs}
\usepackage{dsfont}

\usepackage{hyperref} 
\usepackage{url}      
\usepackage{natbib}

\newcommand{\vc}[1]{{\bf #1 }}

\title{Scaling of Interactive Particle Samplers}

\author{%
  Carlo Albert\thanks{Swiss Federal Institute of Aquatic Science and Technology (Eawag), Switzerland}\\
  \href{mailto:carlo.albert@eawag.ch}{carlo.albert@eawag.ch}%
  \and
  Andreas Scheidegger\footnotemark[1]\\
  \and
  Dmitri Kavetski\thanks{University of Adelaide, Australia}\\
}

\date{}
\begin{document}

\maketitle

\begin{abstract}
    Population Monte Carlo (PMC) samplers are among the most common workhorses for Bayesian inference, offering a range of advantages such as parallelization, reduced autocorrelation and position-dependent jumps. We study the scaling of the effective sample size with population size for Gaussian targets and find a universal curve in the limit of large numbers of particles, providing practical guidance for choosing this number. 
    We also discuss the advantages and disadvantages for two widely used proposal mechanisms (stretch move and differential evolution) for non-Gaussian targets. It turns out that the position-dependence of the stretch move can be an advantage for non-Gaussian targets. However, it also leads to a larger autocorrelation time for Gaussian targets.   
\end{abstract}

\section{Introduction}

Monte Carlo (MC) methods are foundational tools for sampling from probability distributions, particularly in scenarios where analytical solutions are intractable. In Bayesian inference, these methods are pivotal for exploring often complex and analytically inaccessible posterior distributions.

Population Monte Carlo (PMC) algorithms represent an important subclass of MC methods. Unlike traditional approaches such as Importance Sampling or single-chain Markov Chain Monte Carlo (MCMC), PMC algorithms utilize a population of {\em interacting} particles that evolve collectively toward the target distribution. 
Unlike adaptive single-chain Metropolis algorithms (e.g.\cite{haario_2001_adaptiveMetropolis, vihola_2012_RAM}), which rely on a particle's individual history to adjust the jump distribution, PMC methods utilize the population's collective state. 
This population-based approach introduces unique opportunities for enhancing sampling efficiency, including parallel computation, reduced autocorrelation, and better adaptation to the target distribution through position-dependent proposal mechanisms.

The proposal mechanisms of PMC algorithms are often called \emph{moves} and can be classified by the number of partner particles involved:
\begin{enumerate}
    \item \textbf{Single particle:} The Stretch Move (SM) \cite{goodman2010ensemble} displaces a particle by a random distance along the vector defined by a randomly chosen particle.
    \item \textbf{Two particles:} Differential Evolution (DE) \cite{terbraak_2006_DEMC} moves a particle a fixed multiple of the difference between two randomly chosen particles.
    \item \textbf{Multiple partners:} Moves such as the Walk Move \cite{goodman2010ensemble} or the Snooker Move \cite{terbraak2008snooker} employ multiple particles to determine the displacement direction and magnitude.
\end{enumerate}
Contemporary software packages, such as the {\em emcee}-Python package \cite{foreman2019emcee}, integrate these basic moves into more sophisticated algorithms, offering versatile solutions for sampling challenges. 

Despite the advantages of PMC samplers, significant gaps remain in understanding their performance characteristics. For instance, the relative efficacy of different moves is unclear for specific types of target probability density functions (PDFs), such as multimodal distributions or distributions with nonlinear correlations. The impact of the population size on performance is not well-understood, and there are no established guidelines how to determine it, for a given application.

This study aims to address these research gaps through a comprehensive investigation of fundamental PMC moves. Specifically, the objectives are to review the foundational principles of population-based samplers, systematically evaluate their empirical performance across diverse case studies, analyze the strengths and weaknesses of different sampling moves, conduct a scaling analysis of effective sample size as a function of both population size and dimension of the target distribution, and provide practical guidelines for the effective application of PMC algorithms.  We limit ourselves to the two most basic moves requiring just one or two partners, that is the stretch move and the differential evolution move, respectively. 

\section{Basics of Population Samplers}

We describe a population of particles by $d$-dimensional {\em dependent} random variables $\vc X_{\alpha i}$, where $\alpha=1,\dots, L$ denotes the member of the population and $i=1,\dots,N$ is the time-index describing the evolution of the particle. 
We attempt to define algorithms, such that each particle converges to a pre-defined target distribution, $\pi(\vc x)$, i.e., such that each $\vc X_{\alpha i}$ converges to $\pi(\vc x)$ in total variation as $i\rightarrow\infty$.

This is achieved using the {\em Metropolis algorithm} (\cite{metropolis_1953_MRT2}), for each particle update $\vc X_{\alpha i}\rightarrow \vc X_{\alpha,i+1}$, that is, by combining a random jump with a subsequent Metropolis accept/reject step. 
Contrary to adaptive single-chain Metropolis algorithms, the jump distribution is not defined by the history of the chain, but by the positions of the other particles of the population. 
The updating process of the whole population is defined in such a way that the product target distribution $\pi(\vc x_1)\dots\pi(\vc x_L)$ is the {\em stationary} distribution. 
This guarantees that each particle samples the desired target distribution (after the burn-in). It also ensures that the particles remain independent from each other despite the fact that they are interacting with each other. 
To achieve this, we update each particle sequentially, in a random or ordered fashion, with a move $T$ that satisfies the detailed balance condition
\begin{equation}\label{detailedBalance}
    T(\vc x_{\alpha,i+1}|\vc x_{\alpha,i},\{\vc x_{\hat\alpha,i}\})\pi(\vc x_{\alpha,i})
    =
    T(\vc x_{\alpha,i}|\vc x_{\alpha,i+1},\{\vc x_{\hat\alpha,i}\})\pi(\vc x_{\alpha,i+1})
    \,.
\end{equation}
Here, $\{\vc x_{\hat\alpha,i}\}$ is composed of generation $i$ particles (not yet updated) as well as generation $i+1$ particles (already updated) except the $\alpha$-particle.  That is, once a particle has been updated, future particle updates must use its new position. 
It is then straightforward to verify that the product distribution is stationary:
\begin{equation}\label{prodDetailedBalance}
\int T(\vc x_{1,i+1}\dots \vc x_{L,i+1}|\vc x_{1,i}\dots \vc x_{L,i})
\prod_\alpha\pi(\vc x_{\alpha,i})d\vc x_{\alpha,i}
= 
\prod_\alpha\pi(\vc x_{\alpha,i+1})
\,,
\end{equation}
if we set
\begin{equation}\label{eq:propagator}
    T(\vc x_{1,i+1}\dots \vc x_{L,i+1}|\vc x_{1,i}\dots \vc x_{L,i})
=
\prod_{\alpha=1}^L T(\vc x_{\alpha,i+1}|\vc x_{\alpha,i},\{\vc x_{\hat\alpha,i}\})
\,.
\end{equation}
Each single-particle kernel preserves the product target, so their ordered composition does too; the complete ordered sweep need not satisfy detailed balance. A parallel variant splits the population into two subsets and alternately updates one subset using partners only from the other. The simulations here use ordered sequential sweeps. 

We define the two most fundamental moves below.

\subsubsection*{Stretch Move}

For the stretch move, we choose a single partner particle, and jump along the vector the two particles define.
That is, in order to update particle $\vc x_{\alpha,i}$, we randomly choose a partner, $\vc x_{\beta}$, from $\{\vc x_{\hat\alpha,i}\}$ and propose
\begin{equation}
    \vc x^*_{\alpha,i+1}
    =
    \vc x_{\beta}
    +
    z(\vc x_{\alpha,i}-\vc x_{\beta})
    \,,
\end{equation}
where $z$ is drawn from a probability distribution satisfying the symmetry relation $f(1/z)=zf(z)$. A standard choice is $f(z)\propto z^{-1/2}$ restricted to the interval $[1/a,a]$, where $a>1$ is a tuning parameter.
It has recently been pointed out in \cite{chen2025EMCEE} that the excess stretch parameter $a-1$ should scale as $d^{-1/2}$ in order to keep the acceptance rate stable in high dimensions. Following the recommendation in that paper, we set $a=1+2.151/\sqrt{d}$.
This update move is {\em non-homogeneous}, i.e., the direction in which a particle jumps depends on its current position, which can be an advantage for targets with nonlinear correlation structure (see below). 
In order for the detailed balance condition (\ref{detailedBalance}) to be satisfied, we need to replace the standard Metropolis acceptance probability by $\min(1,z^{d-1}\pi(\vc x^*_{\alpha,i+1})/\pi(\vc x_{\alpha,i}))$.
Detailed balance is then readily verified using polar coordinates centered at $\vc x_\beta$ and the symmetry relation for $f(z)$.

\subsubsection*{Differential Evolution}

The differential evolution move uses {\em two} partner particles, $\vc x_\beta$ and $\vc x_\gamma$, chosen as a uniformly sampled ordered pair of distinct particles from $\{\vc x_{\hat\alpha,i}\}$, and proposes
\begin{equation}
    \vc x^*_{\alpha,i+1}
    =
    \vc x_{\alpha,i}
    +
    z(\vc x_\beta - \vc x_\gamma)
    \,.
\end{equation}
Here, the scaling factor $z>0$ is a deterministic tuning parameter. We use the pure difference proposal, without additional additive noise.
Here, we use the standard choice $z=2.38/\sqrt{2d}$.
Contrary to the stretch move, the proposal is homogeneous. While this might be a disadvantage for complex targets, the weaker dependence on the current position reduces the auto-correlation (see below).
Interchanging the two partners reverses the proposal increment with equal probability, so the proposal is symmetric conditional on the other particles. The standard Metropolis acceptance probability therefore satisfies detailed balance.

\subsubsection*{Other MC moves}

Other multiple-partner moves include the Snooker Move (\cite{terbraak2008snooker}) or the Walk Move (\cite{goodman2010ensemble}), which we do not consider in this paper.

\section{Autocorrelation}






At stationarity, the particles $\vc X_{\alpha,i}$ are identically distributed and independent at a fixed time $i$, but different particles can be correlated across time. 
For calculating the loss of effective sample size due to correlations (across time and among chains), we need to introduce a {\em test function}, e.g. by selecting one of the components of vector $\vc X$. To simplify the calculations and w.l.o.g., we choose a test function, $f$, such that $\langle f(\vc X)\rangle =0$ and we denote $\langle (f(\vc X))^2\rangle =\sigma^2$, where the brackets denote the expectation value.

Consider $L$ particles, each of which has chain length $N$, and denote the average of the process $\overline{f(\vc X_i)}:=(1/L)\sum_\alpha f(\vc X_{\alpha i})$.
We want to minimize the fluctuations
\begin{multline}\label{eq:IAT}
    \operatorname{Var}
    \left(
      \frac{1}{NL}\sum_{i,\alpha}f(\vc X_{\alpha i})
    \right)
    =
    \operatorname{Var}
    \left(
        \frac{1}{N}
        \sum_i \overline{f(\vc X_i)}
    \right)
    \\
    =
    \frac{1}{N^2}\sum_{ij} \left\langle \overline{f(\vc X_i)}\,\overline{f(\vc X_j)} \right\rangle
    \approx 
    \frac{1}{N}
    \sum_{i\in \mathbb{Z}} \left\langle \overline{f(\vc X_0)}\,\overline{f(\vc X_i)} \right\rangle
    =:\frac{\tau \sigma^2}{NL}
    \,.
\end{multline}
The factor $\tau$ introduced in the last step defines by how much the total sample size $LN$ is reduced due to correlations of the particles among each other and across time. 
In the second-last step we assume stationarity of the chains and the approximate equal sign means up to terms of order $\mathcal{O}(1/N^2)$.
According to eq. (\ref{eq:IAT}),
\begin{equation}\label{eq:IAT2}
    \tau = 
    \frac{L}{\sigma^2}
    \sum_{i\in \mathbb{Z}}\left \langle \overline{f(\vc X_0)}\,\overline{f(\vc X_i)} \right\rangle
\,.
\end{equation}
Since the particles converge to being {\em independent} for lag zero, i.e., \\
$\langle f(\vc X_{\alpha i})f(\vc X_{\beta i})\rangle =\sigma^2 \delta_{\alpha\beta}$, we have that $\operatorname{Var}(\overline{f(\vc X_i)})=\sigma^2/L$. Hence, $\tau$ coincides with the {\em integrated autocorrelation time (IAT)} of the {\em averaged} process, the performance measure also used by \citep{goodman2010ensemble}.

Because correlations among particles for positive lags do not vanish, $\tau$ is {\em not} the same as the average of the IATs of the individual chains: $\tau\neq\bar{\tau}:=(1/L)\sum_\alpha \tau_{\alpha}$, where $\tau_\alpha$ is the integrated autocorrelation time of chain $\alpha$, because
\begin{equation}
    \tau=
    \frac{1}{L\sigma^2}
    \sum_{i,\alpha,\beta}
    \langle
      f(\vc X_{\alpha 0})
      f(\vc X_{\beta i})
    \rangle
    =        
    \bar\tau+
        (L-1)\kappa
    \,,
\end{equation}
where
\[
 \kappa=\frac{1}{L(L-1)\sigma^2}
 \sum_{i\in\mathbb Z}\sum_{\alpha\ne\beta}
 \langle f(\vc X_{\alpha 0})f(\vc X_{\beta i})\rangle
\]
is the averaged integrated cross-correlation between distinct particles.
The cross-correlation contribution $(L-1)\kappa$ cannot generally be neglected, even when individual pair correlations are small. We therefore estimate the IAT of the population average directly. The estimator has finite-run and truncation uncertainty, as described in Appendix~\ref{compDetails}.

The IAT $\tau$ not only quantifies the effective sample size of a converged chain, but also the convergence time ({\em burn-in}), which is typically assumed to be a multiple of $\tau$.
In this work, we always start from exact samples from the product target distribution and estimate $\tau$ from sufficiently many replica of sufficiently long Markov chains.

\section{Scaling}

Understanding how the effective sample size scales with the number of particles is crucial for deciding how many particles to use in a sampling algorithm. To investigate this, we analyze a $d$-dimensional standard Gaussian target. However, we expect the results to extend to more general, sufficiently regular targets. 

\begin{figure}
    \centering
    \includegraphics[width=0.9\linewidth]{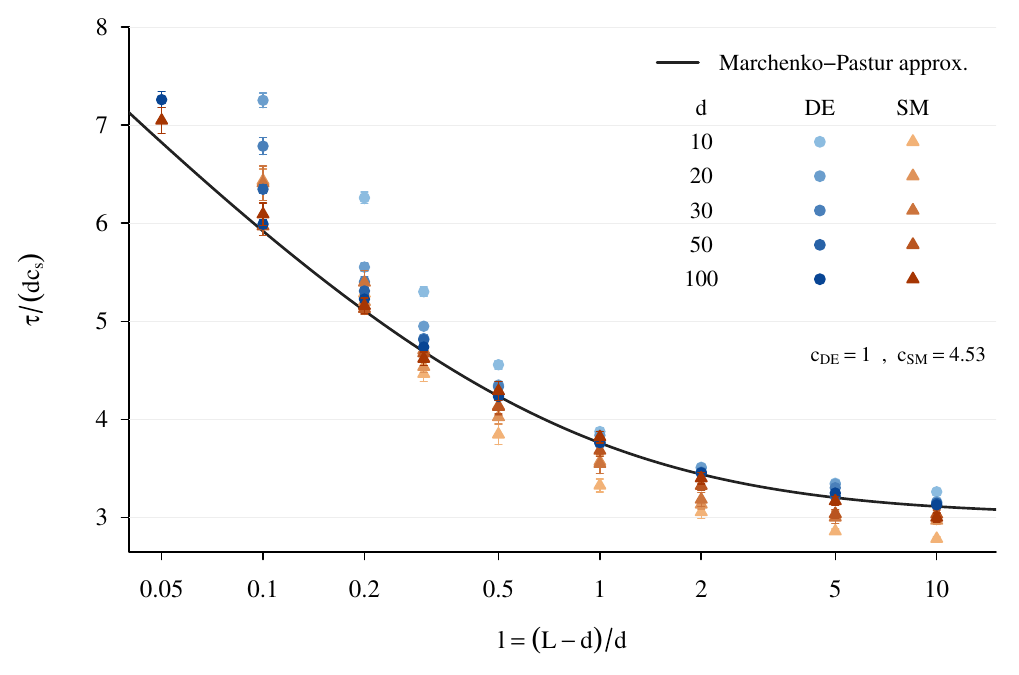}
    \caption{Data collapse according to eq. (\ref{dataCollapse}). On the y-axis are the {\em rescaled} IATs: $\tau/(dc_s)$.}
    \label{fig:dataCollapse}
\end{figure}

We explore the $d$-dimensional standard Gaussian target with $L=d(1+l)$ particles. For Fig.~\ref{fig:dataCollapse} we use $L\ge d+2$, or $l>d^{-1}$.\footnote{For DE, $L=d+1$ preserves the oriented volume of the population simplex and is non-ergodic for the full product target.}
With proposal scales of order $d^{-1/2}$ as described above, we expect the IAT to scale {\em linearly} with $d$ at fixed positive $l$, which is indeed the best we can achieve with {\em any} random-walk Metropolis algorithm. 
In order to estimate the $l$-scaling, we make a {\em mean-field approximation} and assume each particle is interacting with the empirical particle covariance. 
Both the eigenvalues and eigenvectors of the covariance are subject to random fluctuations. However, for large $d$, we can neglect the eigenvalue-fluctuations and derive a {\em universal} curve for the scaling of $\tau$ with $l$ (see Subsect. \ref{sec:universalCurve} below). 
Thus, for $L$ and $d$ sufficiently large, we expect to find the scaling relation of the IAT $\tau$ in terms of dimension ($d$), relative number of walkers ($l$) and move ($s$) to factorise as follows:
\begin{equation}\label{dataCollapse}
    \tau(l,d,s)
    \approx
    d c_s \mathcal{G}(l)\,,
\end{equation}
where $c_s$ is a move-specific {\em clock factor}.
Indeed, Fig. \ref{fig:dataCollapse} shows the data collapse of the rescaled $\tau$'s onto a universal curve $\mathcal{G}(l)$, for sufficiently large $d$.
Once $l$ reaches about $10$, the correlation time has more or less saturated and there is no added value in increasing this number any further (except perhaps for parallelization purposes). 
The smaller the value of $l$, the larger the dimension needs to be for the analytical curve to be a good approximation. 
This is due to the fixed-spectrum assumption we made in Subsect. \ref{sec:universalCurve}, which is rendered invalid if $L-d\sim\mathcal{O}(d^0)$ due to small-eigenvalue fluctuations. 
The relative decorrelation-inefficiency of SM is quantified by the clock factor $c_{SM}\approx 4.53$.
We attribute it to the fact that the particle that is updated is also used for the update {\em vector}.
This is the seemingly unavoidable trade-off that comes with the increased sampler efficiency of SM with complicated posteriors (see next section).

\subsection{Derivation of universal curve}\label{sec:universalCurve}

We use the reversible common-covariance Langevin model of \citet{nusken2019interactingLangevinDiffusion} as a diffusion approximation:
\begin{equation}\label{Langevin}
 d\vc X_\alpha=
 -\gamma_s^2\left(C\vc X_\alpha-\frac{d+1}{L}(\vc X_\alpha-\bar{\vc X})\right)dt
 +\sqrt{2\gamma_s^2 C}\,d\vc W_\alpha,
 \qquad C=\frac1L\sum_\alpha(\vc X_\alpha-\bar{\vc X})(\vc X_\alpha-\bar{\vc X})^T.
\end{equation}
Here $\bar{\vc X}=L^{-1}\sum_\alpha\vc X_\alpha$, and the $\vc W_\alpha$ are independent standard vector Brownian motions, and the state-dependent noise is to be understood according to the It\^o convention. 
The correction is the mobility-divergence drift, since $\operatorname{div}_{X_\alpha}C=(d+1)(\vc X_\alpha-\bar{\vc X})/L$; it ensures invariance of the product Gaussian law. Equations~(\ref{Langevin}) are a model for the collective dynamics, not the exact finite-$L$ diffusion limit of DE or SM. In particular, exact DE partner covariances exclude the particle being updated.
The correction sums to zero, giving
\begin{equation}
 d\bar{\vc X}=-\gamma_s^2 C\bar{\vc X}\,dt
 +\sqrt{2\gamma_s^2 C/L}\,d\vc W_0.
\end{equation}
Define the homogeneous propagator by $\partial_t U(t,0)=-\gamma_s^2 C(t)U(t,0)$ and $U(0,0)=\mathds{1}$. 
Since random fluctuations do not contribute to the correlation with the initial mean, we find that the ACF of the averaged process is given by
\begin{equation}\label{ACFtrU}
 R(t):=\frac{\operatorname{tr}\langle\bar{\vc X}(t)\bar{\vc X}(0)^T\rangle}
 {\langle|\bar{\vc X}(0)|^2\rangle}
 =\frac1d\operatorname{tr}\langle U(t,0)\rangle.
\end{equation}
The brackets include the stationary initial population and the driving noise.

It\^o's product rule for the centered population covariance gives, using eqs. (\ref{Langevin}),
\begin{equation}\label{SDE4C}
 dC=2\gamma_s^2\left(\frac{L+d}{L}C-C^2\right)dt+\sqrt{\frac{2\gamma_s^2}{L}}\sqrt C(dB+dB^T)\sqrt C,
\end{equation}
where $B$ has independent standard Brownian entries. The coefficient of $C$ combines $(d+1)/L$ from the particle drift with $(L-1)/L$ from the quadratic variation of the centered noise. At equilibrium, $LC$ has the Wishart law with $L-1$ degrees of freedom and identity scale.

We diagonalise $C(t)=\sum_a\lambda_a(t)\Pi_a(t)$, where the $\Pi_a(t)$ are rank-one orthogonal projections and the eigenvalues are distinct. Infinitesimal perturbation theory (see, e.g., \citep{kato1966corrected}) gives, in It\^o differential notation,
\begin{equation}\label{eq:IPT}
 d\Pi_a=-[dS,\Pi_a]+\frac12[dS,[dS,\Pi_a]],
 \qquad
 dS=\sum_{a\ne b}\frac{\Pi_a dC\Pi_b}{\lambda_a-\lambda_b}.
\end{equation}
We now hold the eigenvalues fixed and write
\[
 \operatorname{tr}\langle U(t,0)\rangle=\sum_a z_a(t),
 \qquad z_a(t)=\operatorname{tr}\langle\Pi_a(t)U(t,0)\rangle.
\]
Using (\ref{eq:IPT}), omitting time arguments, and writing $dW=dB+dB^T$, we obtain
\begin{align}
 dz_a&=\operatorname{tr}\langle d\Pi_a U+\Pi_a dU\rangle\nonumber\\
 &=\operatorname{tr}\left\langle
 \frac12(dS^2\Pi_a+\Pi_a dS^2-2dS\Pi_a dS)U
 -\gamma_s^2\lambda_a\Pi_a U\,dt\right\rangle\nonumber\\
 &=-\gamma_s^2\lambda_a z_a\,dt
 -\sum_{b\ne a}\frac{\operatorname{tr}\langle
 \Pi_a dC\Pi_b dC\Pi_a U-\Pi_b dC\Pi_a dC\Pi_b U\rangle}
 {(\lambda_a-\lambda_b)^2}\nonumber\\
 &=-\gamma_s^2\lambda_a z_a\,dt
 -\sum_{b\ne a}\frac{2\gamma_s^2\lambda_a\lambda_b}{L(\lambda_a-\lambda_b)^2}
 \operatorname{tr}\langle\Pi_a dW\Pi_b dW\Pi_a U-\Pi_b dW\Pi_a dW\Pi_b U\rangle\nonumber\\
 &=\left[-\gamma_s^2\lambda_a z_a-\sum_{b\ne a}r_{ab}(z_a-z_b)\right]dt,
 \qquad z_a(0)=1,\label{eq:unravel}
\end{align}
where $r_{ab}=4\gamma_s^2\lambda_a\lambda_b/[L(\lambda_a-\lambda_b)^2]$.
In the second line, we used $dU=-\gamma_s^2CU\,dt$. In the fourth line, we used (\ref{SDE4C}) and the fact that its drift commutes with the projections. Finally, in the instantaneous eigenbasis, $\langle(dW_{ab})^2\rangle=2dt$ for $a\ne b$, which gives the factor $4$ in $r_{ab}$.
Thus angular averaging can be seen as a jump process on the fixed spectrum, with jump rates $r_{ab}$ and killing rates $\gamma_s^2\lambda_a$, following the eigenvector moment-flow idea of \citet{bourgade2017eigenvectorMomentFlow}. If the eigenvalues fluctuate, the rates and $\lambda_a$ remain inside the expectation and the ODE does not close.

At fixed $q=d/L<1$, the stationary empirical spectrum converges as $d,L\to\infty$ to the {\em Marchenko--Pastur} density
\begin{equation}
 \rho_{MP}(\lambda)=\frac{\sqrt{(\lambda_+-\lambda)(\lambda-\lambda_-)}}{2\pi q\lambda},
 \qquad \lambda_\pm=(1\pm\sqrt q)^2,
 \qquad \Lambda=[\lambda_-,\lambda_+].
\end{equation}
The vector $\vc z$ then turns into a continuous function, $f(\lambda)$, on $\Lambda$. Its dynamics is derived from eq. (\ref{eq:unravel}) and described by the Hamiltonian on the Hilbert space $L^2(\Lambda,\rho_{MP})$:
\begin{equation}
 (Hf)(\lambda)=\gamma_s^2\lambda f(\lambda)
 +4\gamma_s^2q\operatorname{PV}\int_\Lambda
 \frac{\lambda\eta}{(\lambda-\eta)^2}[f(\lambda)-f(\eta)]\rho_{MP}(\eta)d\eta.
\end{equation}
For this continuum surrogate, equation~(\ref{eq:unravel}) has initial function $1$ and gives
\begin{equation}\label{ACFH}
 R(t)=\langle1,e^{-tH}1\rangle
 =\operatorname{tr}_{L^2(\Lambda,\rho_{MP})}\left(f_0e^{-tH}\right),
 \qquad f_0=|1\rangle\langle1|.
\end{equation}
The two-sided discrete IAT is then calculated as
\begin{equation}
 \tau=\langle1,\coth(H/2)1\rangle
 \simeq 2\langle1,H^{-1}1\rangle.
\end{equation}
The last expression is the continuous-time approximation on the slow diffusive scale. We evaluate it by quadrature and a linear solve, as detailed in Appendix~\ref{compDetails}.
For the reference DE clock we use
$\gamma_{DE}^2=2.38^2\Phi(-2.38/2)/d\simeq0.662866/d$, where $\Phi$ is the standard normal distribution function. This yields the curve $\mathcal G(l)=2\langle1,\widehat H^{-1}1\rangle/0.662866$, with $\widehat H=H/\gamma_s^2$. We set $c_{DE}=1$ and calibrate the single relative SM clock empirically. Finite-population effects and the approximation of discrete moves by (\ref{Langevin}) are not absorbed exactly by this scalar rescaling.

\section{Non-Gaussian targets}

Realistic sampling targets are often very far from being Gaussian, and exhibit features such as multi-modality or non-linear correlations. 
In this chapter, we assess how the two basic moves handle such complications. 
As test cases exemplifying these features, we use a bimodal and a Rosenbrock distribution, respectively.

\subsubsection*{Bimodal}

A bimodal mixture of Gaussian distributions is used as a prototypical example of a multimodal distribution:
\begin{equation}
\label{equ:bimodalPdfEq}
    \pi_B(\vc x;\boldsymbol{\mu}_1,\sigma_1,\boldsymbol{\mu}_2,\sigma_2,w)=
    \frac{w}{2\pi \sigma_1^2} \exp\left( -\frac{1}{2\sigma_1^2}||\vc x-\boldsymbol{\mu}_1||^2\right) 
   +
     \frac{1-w}{2\pi \sigma_2^2} \exp\left( -\frac{1}{2\sigma_2^2}||\vc x-\boldsymbol{\mu}_2||^2\right)
     \,.
\end{equation}
Here we limit our attention to 2-dimensional problems.
An exact sampler for the bimodal Gaussian distribution is given by first sampling a uniform random number to select the component followed by sampling from the Gaussian distribution with the component-specific parameters.
We set $w=0.8$, $\boldsymbol\mu_1=(-4,0)$, $\boldsymbol\mu_2=(4,0)$, $\sigma_1=\sigma_2=1$, and $L=20$. The observable is the first coordinate, aligned with the mode separation. 

\begin{figure}
\centering
\includegraphics[width=\textwidth]{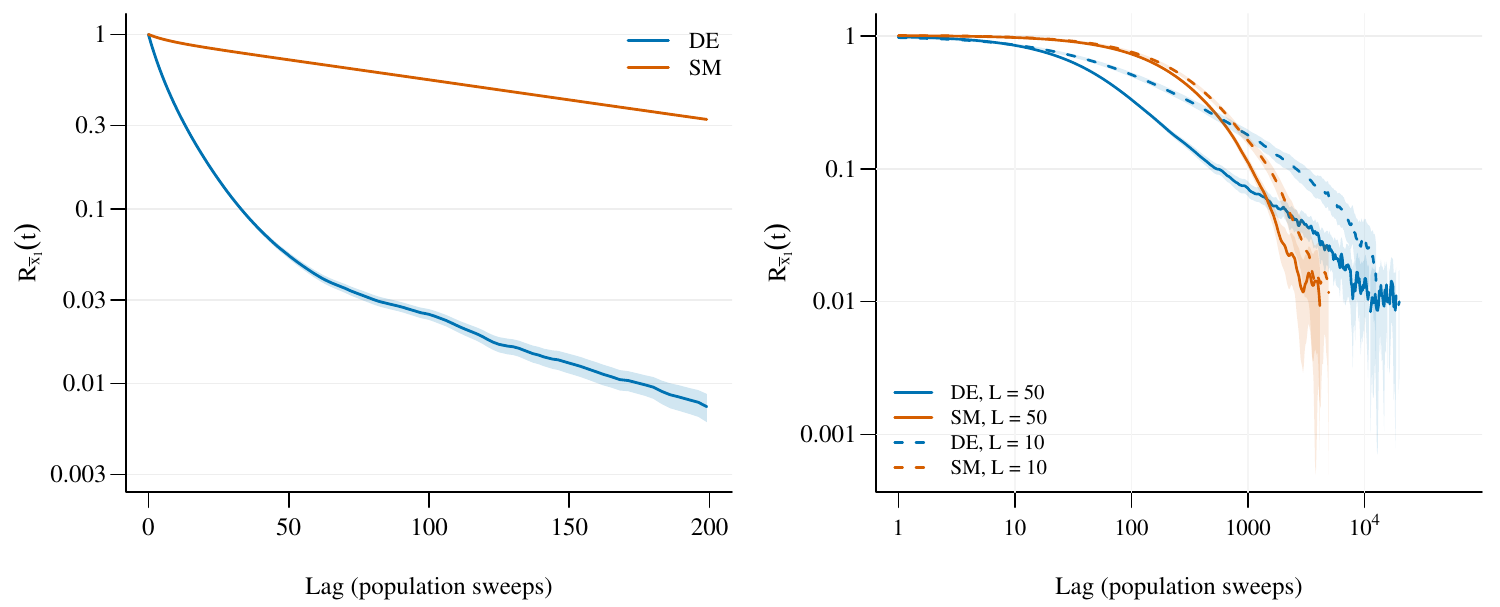}
\caption{Autocorrelation of the first-coordinate population mean for two-dimensional non-Gaussian targets. Left: bimodal Gaussian mixture with masses $0.8$ and $0.2$, centers $(-4,0)$ and $(4,0)$, unit component covariances, and $L=20$; DE is blue and SM orange throughout. Right: Rosenbrock target with $\mu_1=\sigma_1=1$ and $\sigma_2=0.1$; solid curves show $L=50$ and dashed curves $L=10$. Numerical details can be found in the Appendix.}
\label{fig:nonGaussian}
\end{figure}

The ACF exhibits more than one relaxation scale, consistent with motion within the components and redistribution between them (Fig.~\ref{fig:nonGaussian}). At the chosen tunings DE decorrelates much faster than SM. In the case of DE (with the standard choice $z=2.38/\sqrt{4}\approx 1.19$), selecting the two partner particles from different modes in the right order (which is expected to happen for the fraction $w(1-w)$ of the cases), the particle will jump to the other mode. In the case of SM, we have that $\vc x^*_\alpha=\vc x_\alpha+(z-1)(\vc x_\alpha-\vc x_\beta)$.
In order for the modes to mix, the partner particle $\vc x_\beta$ needs to be selected from the other mode (which happens for a fraction of $w$ (or $1-w$) of the cases). 
In two dimensions, the effective jump width $z-1$ fluctuates between $1/a-1$ and $a-1$, with $a=1+2.151/\sqrt{2}\approx 2.52$, which makes mixing of the modes much less likely.  
The small-amplitude slow DE contribution is associated with occasional population mode collapse and subsequent rediscovery of the other component.

\subsubsection*{Rosenbrock}

The Rosenbrock function for $d=2$ is used as a prototypical example of a distribution with non-linear correlation, i.e. a distribution with a nonlinear dependence between its variables.
Its density reads as 
\begin{equation}
\label{equ:rbPdfEq}
    \pi_R(x_1,x_2;\mu_1,\sigma_1,\sigma_2) = \frac{1}{2\pi\sigma_1\sigma_2} \exp\left(- \frac{(x_1-\mu_1)^2}{2\sigma_1^2} - \frac{(x_2-x_1^2)^2}{2\sigma_2^2}\right)\,.
\end{equation}
An exact sampler for the Rosenbrock distribution is given by sampling $x_1$ from a Gaussian distribution with mean $\mu_1$ and standard deviation $\sigma_1$ followed by sampling $x_2$ from a Gaussian distribution with mean $x_1^2$ and standard deviation $\sigma_2$.
We use $\mu_1=\sigma_1=1$, $\sigma_2=0.1$, and $L=10,50$. The curved ridge can lead to very low acceptance for proposals that move far from the local high-density region. In Fig.~\ref{fig:nonGaussian}, DE relaxes faster initially, whereas SM has smaller ACFs at the resolved longer lags, with the difference more pronounced for the smaller population. 
The slower initial decay is outweighed by the steeper tails and gives SM an advantage in terms of the IAT. 
The position-dependent SM direction is a plausible explanation for this behavior. 

\section{Conclusions}

For PMC, the uncertainty of a pooled sample average is governed by the autocorrelation of the population-averaged observable. Although particles are independent at a common time under the product target, cross-particle correlations at nonzero lags contribute to the effective sample size. Averaging individual-particle IATs generally omits these contributions, making the population-mean IAT the appropriate quantity for the comparisons reported here.

For Gaussian targets, the simulations support an approximate collapse of the IAT after rescaling by dimension and a move-specific clock. A common-covariance diffusion model, followed by a fixed-spectrum Marchenko--Pastur approximation, provides a useful prediction for the dependence on relative population size. The agreement improves away from the finite-dimensional hard edge, while fluctuations of the smallest eigenvalues and move-dependent finite-dimensional effects limit the approximation near the minimum population size. Increasing the population eventually gives diminishing reductions in IAT per sweep; practical choices should also account for the number of proposals and the available parallelism.

The comparison between DE and SM depends on target geometry, tuning, and the observable. DE is faster for the Gaussian and bimodal examples studied here, whereas SM has smaller long-lag correlations for the narrow Rosenbrock target despite its slower initial relaxation. These results suggest mixing of the two moves, whose complementary proposal geometries may be useful in practice.

\appendix

\section{Computational Details}\label{compDetails}

\subsection{Simulation protocol}

All data in Figs.~\ref{fig:dataCollapse} and \ref{fig:nonGaussian} were generated using the original discrete Metropolis samplers, not a numerical integration of the diffusion model. One population update, or sweep, is an ordered pass through all $L$ particles, with exactly $L$ proposals including rejections. Each proposal uses the current population, including partners updated earlier in that sweep. DE chooses two distinct ordered partners uniformly, excluding the updated particle, and uses $z=2.38/\sqrt{2d}$ without additive noise. SM chooses one other particle uniformly, draws its stretch factor with density proportional to $z^{-1/2}$ on $[1/a,a]$, and uses $a=1+2.151/\sqrt d$ with the acceptance factor $z^{d-1}$.

Each replica starts with independent exact draws from the target, so no burn-in is discarded. Replicas use separate random seeds. Simulations use double-precision arithmetic, with population means saved in single precision after every complete sweep and no thinning; the ACF calculation uses double precision. For Fig.~\ref{fig:dataCollapse}, all $d$ coordinates of the population mean are recorded. Figure~\ref{fig:nonGaussian} uses its first coordinate. Replicas, rather than coordinates or individual particles, are the independent units for uncertainty estimation.

\subsection{Autocorrelation and integrated autocorrelation estimates}

Let $m^{(r)}_j(n)=L^{-1}\sum_\alpha X^{(r)}_{\alpha j}(n)$ be the population mean of coordinate $j$ in replica $r$ after sweep $n$, and let $\mu_j$ and $v_j$ be the exact target coordinate mean and variance. For a replica of length $N$, we estimate
\begin{equation}\label{eq:acfEstimator}
 \widehat R^{(r)}_j(k)=\frac{L}{v_j(N-k)}
 \sum_{n=1}^{N-k}[m^{(r)}_j(n)-\mu_j]
 [m^{(r)}_j(n+k)-\mu_j],\qquad 0\le k<N.
\end{equation}
Lag products are evaluated by FFT with zero padding to a power of two of length at least $2N-1$, followed by division by the actual number of time pairs $N-k$. We subtract the known target mean, not the empirical trajectory mean, and normalize by the known population-mean variance $v_j/L$, not the observed lag-zero covariance. Consequently, the estimated lag-zero ACF need not be exactly one.

For the standard Gaussian target, $\mu_j=0$ and $v_j=1$. We first average the coordinate ACFs within each replica,
$\widehat R^{(r)}(k)=d^{-1}\sum_j\widehat R^{(r)}_j(k)$, and then average across the $n_{\rm rep}$ replicas. For Fig.~\ref{fig:nonGaussian}, we use only $j=1$, with $(\mu_1,v_1)=(-2.4,11.24)$ for the mixture and $(1,1)$ for the Rosenbrock target. At every lag the standard error is the sample standard deviation across replica estimates divided by $\sqrt{n_{\rm rep}}$.

The Fig.~\ref{fig:dataCollapse} IAT estimates use the two-sided discrete convention,
\begin{equation}
 \widehat\tau(W)=1+2\sum_{k=1}^{W}\widehat R(k).
\end{equation}
We truncate the ACF sum at the smallest lag \(W\) that is at least sixteen times the resulting positive IAT estimate. We assess sensitivity to this choice by varying the truncation rule. The same selected window is used for all replica contributions. Error bars are one standard error of the replica integrals $1+2\sum_{k=1}^{W}\widehat R^{(r)}(k)$, retaining their within-replica lag and coordinate correlations. Window factors $4,6,8,10,12,16,20$ and first-half versus second-half diagnostics are retained for comparison. These finite-window errors do not quantify omitted tail area or the uncertainty in selecting the window.

The bands in Fig.~\ref{fig:nonGaussian} are pointwise estimates $\widehat R(k)\pm1.96\,\mathrm{SE}(k)$, not simultaneous confidence bands. ACFs were calculated through lag $N/4$; the bimodal panel displays lags $0$--$200$. For the logarithmic Rosenbrock panel, points with $\widehat R\le2\,\mathrm{SE}$ are omitted, and a curve is stopped after six consecutive unresolved points on the plotting grid. Signed estimates are retained in the analysis data. Display truncation does not establish that the remaining tail is zero.

\subsection{Gaussian run counts and clock calibration}

Table~\ref{tab:gaussianRuns} lists every $(d,L)$ setting used in the 78-point Fig.~\ref{fig:dataCollapse}; each listed $L$ is a separate setting for each move. The gap is $l=(L-d)/d$. Counts refer to the selected production trajectories, excluding pilots and superseded runs.

\begin{table}[htbp]
\centering
\small
\begin{tabular}{rlrrrr}
\toprule
& & \multicolumn{2}{c}{DE} & \multicolumn{2}{c}{SM}\\
$d$ & $L$ & $N$ & $n_{\rm rep}$ & $N$ & $n_{\rm rep}$\\
\midrule
10 & 12, 13 & 65536 & 64 & 65536 & 64\\
10 & 15, 20, 30 & 32768 & 64 & 32768 & 64\\
10 & 60, 110 & 32768 & 64 & 131072 & 128\\
20 & 22, 24, 26 & 131072 & 64 & 131072 & 64\\
20 & 30, 40, 60, 120, 220 & 65536 & 64 & 65536 & 64\\
30 & 33, 36, 39 & 65536 & 64 & 65536 & 64\\
30 & 45, 60, 90, 180 & 32768 & 64 & 32768 & 64\\
50 & 55, 60, 65 & 131072 & 64 & 131072 & 64\\
50 & 75, 100, 150, 300, 550 & 65536 & 64 & 65536 & 64\\
100 & 105, 110, 120, 130 & 131072 & 64 & 131072 & 64\\
100 & 150, 200, 300, 600, 1100 & 65536 & 64 & 65536 & 64\\
\bottomrule
\end{tabular}
\caption{Population sweeps $N$ per replica and independent replica counts for Fig.~\ref{fig:dataCollapse}. Every population size in a row uses the stated counts.}\label{tab:gaussianRuns}
\end{table}

The selected Gaussian runs comprise 5120 replica trajectories, 402653184 population sweeps, and 51711574016 individual-particle proposals. For SM at $(d,L)=(10,60)$ and $(10,110)$, only the longer 128-replica runs are used; they are not pooled with the earlier shorter runs.

We fix $c_{DE}=1$. The common SM clock is the inverse-variance weighted mean of the two matched SM/DE IAT ratios at $d=100$ and $l=1,2$, giving $c_{SM}=4.5303\pm0.0567$ (one standard error). It is then held fixed for every plotted dimension and gap, including the $l=5,10$ validation points. The ordinate is $\widehat\tau/(dc_s)$. Individual error bars do not include the shared multiplicative uncertainty in $c_{SM}$.

\subsection{Non-Gaussian run counts}

For the bimodal panel, each move has 128 replicas of 262144 sweeps with $L=20$. Initial particles are drawn independently by choosing a component with probabilities $0.8,0.2$ and then drawing from its unit-covariance Gaussian at $(-4,0)$ or $(4,0)$.

For the Rosenbrock panel, each of the four move/population settings (DE or SM, $L=10$ or $50$) has 64 replicas of 262144 sweeps, with $\mu_1=\sigma_1=1$ and $\sigma_2=0.1$. Exact initial draws use independent standard normals $Z_1,Z_2$ and set $X_1=1+Z_1$, $X_2=X_1^2+0.1Z_2$. These are stationary ACF experiments, not burn-in experiments.

\subsection{Numerical evaluation of the analytical curve}

We factor $H=\gamma_s^2\widehat H$ and solve $\widehat H u=1$. To integrate against the Marchenko--Pastur density, we use
$\lambda(\theta)=1+q-2\sqrt q\cos\theta$, $0\le\theta\le\pi$, for which
$\rho_{MP}(\lambda)d\lambda=2\sin^2\theta/(\pi\lambda(\theta))\,d\theta$.
Gauss--Legendre quadrature with 1024 nodes gives normalized positive weights $w_i$. The discretized operator is
\begin{equation}
 (\widehat H_{n_q} u)_i=\lambda_i u_i+
 4q\sum_{j\ne i}w_j\frac{\lambda_i\lambda_j}{(\lambda_i-\lambda_j)^2}(u_i-u_j).
\end{equation}
Here $n_q=1024$ is the quadrature size. We symmetrize with the square roots of the weights and use a Cholesky solve. The integral is $\Theta\simeq\sum_iw_iu_i$, and the plotted curve is $\mathcal G(l)=2\Theta/0.662866$ with $q=1/(1+l)$. Checks using 512, 1024, and 2048 nodes were performed at $l=0.03,0.05,0.1,0.2,0.3,0.5,1,2,3,5,10,50$; the 1024-to-2048-node changes are below $0.10\%$ at these points. This checks quadrature resolution, not the accuracy of the diffusion or fixed-spectrum approximations.

The simulation, analysis, and plotting scripts, selected run counts, and retained diagnostics are available in the accompanying repository, \url{https://github.com/carloalbert/EMCEE}. Julia is used for simulation and the operator calculation, Python/NumPy for FFT analysis, and R for the figures. Raw replica traces are retained separately from the Git-tracked summaries.

\bibliographystyle{apalike}
\bibliography{refs,refs_corrections}

\end{document}